\documentclass{PoS}

\title{Anomaly Inflow and Membranes in QCD Vacuum}

\ShortTitle{Anomaly Inflow and Membranes in QCD Vacuum}

\author{H. B. Thacker\\
Department of Physics \\
        University of Virginia, USA\\
        E-mail: \email{hbt8r@virginia.edu}}

\author{\speaker{Chi Xiong}
\\
        Institute of Advanced Studies, Singapore\\
        Nanyang Technological University \\
        E-mail: \email{xiongchi@ntu.edu.sg}}

\abstract{
We study the membrane-like structure of topological charge density and its fluctuations in the QCD vacuum. Quark zero modes are localized on the membranes and the resultant gauge anomaly is cancelled by the gauge variation of a Chern-Simons type effective action in the bulk via the anomaly inflow mechanism. The coupling between brane fluctuations, described by the rotations of its normal vector, and the Chern-Simons current provides the needed anomaly inflow to the membrane. This coupling is also related to the axial U(1) anomaly which can induce brane punctures, and consequently quark-antiquark annihilation across the brane. As the Chern-Simons current has a long-range character, together with membranes it might lead to a solution to the confinement problem. 
          }

\FullConference{Xth Quark Confinement and the Hadron Spectrum\\
                 8-12 October 2012\\
                 TUM Campus Garching, Munich, Germany}

\begin{document}

\section{Introduction}

Topological charges and their distributions play important roles in non-perturbative quantum chromodynamics (QCD). Different topological defects in the QCD vacuum have been studied by large-$N_c$ methods, holographic QCD and lattice simulations. It is suggested that codimension-one membranes or domain walls exist in the QCD vacuum by theoretical studies \cite{Witten79, Witten98} and Monte Carlo simulations \cite{Horvath03, Horvath05, Ilgenfritz07, Ilgenfritz08, Hynes08}. These membrane-like structures can be considered as ``I2-branes" which are the intersections of the D6-brane and D4-branes from the holographic QCD point of view, and the QCD $\theta$ parameter comes from the Wilson line of the closed string Ramond-Ramond U(1) gauge field \cite{Witten98, Thacker10, TX1}. This also supports the previous large-$N_c$ arguments about the existence of multiple metastable $k$ vacua, separated by the membranes across which $\theta$ jumps by $2\pi$. A recent lattice study with overlap fermions \cite{Buividovich11} shows that cooling gauge configurations changes the Hausdorff dimension of regions where the topological charge is localized gradually from $d = 2 \sim 3$ to the total space dimension ($d=4$), and chiral fermion zero modes are localized on
structures with fractal dimension $d = 2 \sim 3$, which favors the vortex or domain-wall nature of localization.     

The localization of chiral fermion zero modes on vortices or domain-walls brings an interesting topic --- anomaly inflow mechanism \cite{Callan-Harvey, Green} into the study of topological charge distribution in the QCD vacuum.
It has been shown \cite{Callan-Harvey} that chiral fermion zero modes are localized on an axion string or a domain-wall embedded in a higher dimensional spacetime. When coupled to some external gauge potential, gauge anomaly appears on the string or domain-wall and it is cancelled by the gauge variation of an effective action, a Chern-Simons type coupling living in the higher dimensional spacetime. In Ref.\cite{TX1} we applied the intersecting brane version of the anomaly inflow mechanism \cite{Green} to the D6-D4 brane system. In Ref. \cite{TX2} and here we study the anomaly inflow in the QCD vacuum without reference to the higher dimensional string theory. The Ramond-Ramond U(1) gauge field appears as an auxiliary field, describing the membrane-like excitations in four-dimensional QCD. We found that the analogue of the Ramond-Ramond U(1) gauge field, $\partial_{\mu} \theta$ and the Chern-Simons current $K_{\mu}$, play crucial roles in the studies of QCD vacuum structure, chiral symmetry breaking, and probably confinement as well.

\section{Geometric Preparations}

We start with a pure geometric description of codimension-one membranes embedded in four-dimensional spacetime. Neglecting its thickness and idealizing it as a three-dimensional surfaces, we parametrize
the position of a membrane by embedding functions 
\begin{equation}
x^{\mu} = x^{\mu} (y_i), ~ i=0,1,2.
\end{equation}
where $y^i~(i=0,1,2)$ are the brane coordinates. In general one can define an induced metric on the brane,
\begin{equation}
h_{ij} = g_{\mu\nu} \partial_{i} x^{\mu} \partial_{j} x^{\nu}
\end{equation}
which describes a curved three-dimensional brane embedded in a flat four-dimensional spacetime.  
One can take a ``static gauge"
\begin{equation}
x^{i} = y^i,~ (i=0,1,2), ~~ x^3 = \phi
\end{equation}
The induced metric then becomes
\begin{equation} \label{braneMetric}
h_{ij} = g_{ij} + \partial_{i} \phi \partial_{j} \phi, ~~ (i,j = 0,1,2).
\end{equation}
which can be used to build a Nambu-Goto type effective action to describe the membrane dynamics. However we are only interested in the coupling between the brane fluctuation and some gauge potential, which will be linear in $\partial_{i} \phi$ (or linear in $\partial_{\mu} \theta$ as it will be shown later). We also ignore the translation and acceleration of the membrane and focus on rotations or other motions which only change the orientation of the branes. (Note that in multiple-brane cases dilatory collective modes of brane motion may be important). Embedding conditions require that \cite{Clark}
\begin{equation} \label{embedding}
g_{\mu\nu} \partial_i x^{\mu} n^{\nu} = 0, ~~~~~~g_{\mu\nu} n^{\mu} n^{\nu} = 1,
\end{equation}
which are the orthonormal condition of the tangent vector $ \partial_i x^{\mu} $ and the normal vector $n^{\mu}$ of the membrane, and the normalization of $ n^{\mu} $ respectively. These two equations can be solved for $n_{\mu} = (n_i, n_3) $
\begin{equation} \label{nvector}
n_i = - \frac{\partial_i \phi}{\sqrt{1 + \partial^k \phi \partial_k \phi }}, ~~n_3 =  \frac{1}{\sqrt{1 + \partial^k \phi \partial_k \phi }}
\end{equation}
As the $\theta$ parameter jumps by $2\pi$ across the membrane, we need a step function $ \theta(t) $
\begin{equation}
\theta (t) = \left\{ \begin{array}{ll}
1, & ~t \geqslant 0 \\
0, & ~t < 0 
\end{array} \right.
\end{equation}
and its generalization. For describing a membrane with a jumping $ \theta $ term, we choose
\begin{equation}
t = \Sigma (x^0, x^1, x^2, x^3) 
\end{equation}
where $ \Sigma(x^0, x^1, x^2, x^3) =0$ is the equation of a three dimensional surface (or location of the membrane).  The step function then becomes
\begin{equation}
\theta (\Sigma(x^0, x^1, x^2, x^3)) = \left\{ \begin{array}{ll}
1, & ~\Sigma(x^0, x^1, x^2, x^3) \geqslant 0 \\
0, & ~\Sigma(x^0, x^1, x^2, x^3) < 0 
\end{array} \right.
\end{equation}
Taking derivative with respect to the spacetime coordinates $x^{\mu}$ yields
\begin{equation}
\partial_{\mu} \theta = \delta (\Sigma) ~\partial_{\mu} \Sigma 
\end{equation}
Noticing that the vector $ \partial_{\mu} \Sigma $  is perpendicular to the membrane (but not normalized), we should have
\begin{equation} \label{dtheta}
\partial_{\mu} \theta \propto \delta (\Sigma) ~n_{\mu}.  
\end{equation}
This can be worked out more explicitly. Suppose the equation of the membrane $\Sigma(x^0, x^1, x^2, x^3) =0$ can be solved by
\begin{equation}
x^3 = \phi (x^0, x^1, x^2)
\end{equation} 
%
%so one may write the equation of the hypersurface as
%
%\begin{equation}
%\Sigma(x^0, x^1, x^2, x^3) = x^3 - \phi (x^0, x^1, x^2) = 0
%\end{equation}
%
and the gradient of the function $\Sigma$ gives a vector
\begin{equation}
\partial_{\mu} \Sigma = (-\partial_i \phi, 1) ~\propto~ n_{\mu}
\end{equation}
which is the same as the normal vector $n_{\mu}$ in (\ref{nvector}) up to normalization. From eqt. (\ref{dtheta}) we see that $ \partial_{\mu} \theta $ is proportional to the product of a delta function and the normal vector of the membrane.

%%%%%%%%%%%%%%%%%%%%%%%%%%%%%%%%%%%%%%%%%%%%%%%%%%%%%%%%%%%%%%%%%%%%%%%%%%%%%%
%
\section{Anomaly Inflow and Axion String}
%
%%%%%%%%%%%%%%%%%%%%%%%%%%%%%%%%%%%%%%%%%%%%%%%%%%%%%%%%%%%%%%%%%%%%%%%%%%%%%%

Having identified the membrane ``inflow current" $\partial_{\mu} \theta \propto \delta (\Sigma) ~n_{\mu}$,
one may easily write its coupling to the Chern-Simons current $K_{\mu}$
\begin{equation} \label{CSphi}
\mathcal{L_{\textrm{\tiny{int}}}} = \partial_{\mu}\theta \, K^{\mu}
\end{equation}
where the Chern-Simons current
\begin{equation}
K_{\mu} = \epsilon_{\mu\alpha\beta\gamma} \textrm{Tr} (A^{\alpha} \partial^{\beta} A^{\gamma} + \frac{2}{3} A^{\alpha} A^{\beta} A^{\gamma})
\end{equation}
comes from the Chern-Simons form $ \mathcal{K}_{cs} = \textrm{Tr} (A \wedge F - \frac{1}{3} A \wedge A \wedge A ) $.
The new interaction (\ref{CSphi}) is manifestly Lorentz invariant. Integrated by parts it gives a term resembling the QCD $\theta$-term 
\begin{equation} \label{FF}
\theta \, F_{\mu\nu} \tilde{F}^{\mu\nu}
\end{equation}
with the help of the identity (using differential forms)
\begin{equation}
d\, \mathcal{K}_{cs} = \textrm{Tr} ~F \wedge F.
\end{equation}
Note that the topological charge density (\ref{FF}) is gauge invariant, while the current-current coupling (\ref{CSphi}) is not. This is because the Chern-Simons form transforms under the gauge transformation
\begin{eqnarray}
A &\longrightarrow & A' = g^{-1} A g + g^{-1} dg, \cr
\mathcal{K}_{cs} & \longrightarrow & \mathcal{K}'_{cs} = \mathcal{K}_{cs} - d\,\textrm{Tr}(dg g^{-1} A ) - \frac{1}{3} \textrm{Tr} [ (g^{-1} dg)^3]. 
\end{eqnarray}
The gauge variation of (\ref{CSphi}) is crucial in the anomaly-inflow mechanism. 
It can cancel the gauge anomaly due to the fermion zero modes localized at lower dimensional topological defects \cite{Callan-Harvey, Green}. Let us take Callan and Harvey's axion-string model as an example --- an axion string, described by a complex scalar field $\Phi = f(\rho) e^{i \vartheta}$, is embedded in a four-dimensional bulk space. Fermions coupled to the string have chiral zero modes localized on it. If the fermions are also coupled to some gauge potential, there is a two-dimensional gauge anomaly due to the fermion zero modes
\begin{equation}
D^k J_k = \frac{1}{2\pi} \epsilon^{ij} \partial_i A_j, ~~~~~~i,j,k={0,3}
\end{equation}
which can only be cancelled by a Chern-Simons coupling similar to (\ref{CSphi}) \cite{Callan-Harvey}
\begin{equation} \label{axionstring}
S_{\textrm{\tiny{eff}}} = - \frac{1}{8 \pi^2} \int d^4x \, \partial_{\mu} \vartheta \, K^{\mu} 
\end{equation}   
where $\vartheta$ is the phase angle of the axion field $\Phi$. Integrating by parts puts (\ref{axionstring}) into the form of an axial anomaly in four dimensions. The gauge variation of the (\ref{axionstring}) leads to the descent equation \cite{Zumino83, Stora83}
\begin{eqnarray} \label{descent}
\delta \int_{M} d\vartheta \wedge \mathcal{K}^0_{3} &=& \int_{M} d\vartheta \wedge d \mathcal{K}^1_{2} \cr
&=& - \int_{M} d^2\vartheta \wedge  \mathcal{K}^1_{2}
\end{eqnarray}
which relates the four dimensional axial anomaly $d \mathcal{K}^0_{3} $ and the two dimensional gauge anomaly $ \mathcal{K}^1_{2} $.
Note that the phase angle $\vartheta$ is ambiguous at the origin and hence $d^2\vartheta$ is singular \cite{Callan-Harvey}
\begin{equation} \label{deltaxy}
d^2\vartheta = 2 \pi \delta(x) \delta(y) dx \wedge dy.
\end{equation}

\section{Anomaly-inflow and Membrane Dynamics}

In Ref.\cite{Callan-Harvey} the topological defects are static and the $\theta$-variable should be promoted to a dynamical field when fluctuations of the topological defects are included. The anomaly inflow mechanism can still be applied but the dynamical field, e.g. the Ramond-Ramond potential in Refs.\cite{Green, TX1} should vary under the gauge transformation, as dictated by the descent equations \cite{Zumino83, Stora83}. In Ref.\cite{TX1} we studied the anomaly inflow for the intersecting branes (D6-D4 "I-brane") which requires
\begin{equation} \label{dc7}
\delta(d C_7) = \mu \delta \mathcal{K}_{cs} \wedge \delta_{D4}
\end{equation}
where the Ramond-Ramond potential $C_7$ couples to the D6  brane and $\delta_{D4}$ is a 5-form delta-function distribution transverse to the D4 brane, similar to (\ref{deltaxy}) and $\mu$ is related to the brane tension.
Here we restrict ourselves to four dimensions and consider the Ramond-Ramond field as an auxiliary field describing membrane-like excitations of the Yang-Mills field. As shown in eqt. (\ref{dtheta}) in the previous section, for a fluctuating or curved membrane $\partial_{\mu} \theta$ is proportional to the normal vector of the surface. Eqt.(\ref{dc7}) suggests that
\begin{equation} \label{2DAI}
\delta (\partial_{\mu} \theta ) = - \delta K_{\mu}.
\end{equation}
Now let us illustrate the formulation of the membrane dynamics by a 2-dimensional $U(1)$ gauge theory --- the Schwinger model \cite{Kogut75, Coleman75, TX2}. In this case, the Chern-Simons current is 
\begin{equation}
K_{\mu}= \varepsilon_{\mu\nu}A^{\nu}.
\end{equation}
Note that in two dimensions $K_{\mu}$ is always "perpendicular" to the gauge potential $A_{\mu}$ since 
$K_{\mu} A^{\mu} = 0$.   
$A_{\mu}$ can be decomposed as
\begin{equation}  \label{2dA}
A_{\mu}= \varepsilon_{\mu\nu}\partial^{\nu}\sigma + \partial_{\mu}\Omega  
\end{equation}
which can be thought as a gauge potential in two-dimensional Lorentz gauge plus an arbitrary gauge transformation with parameter $\Omega$. This induces a gauge variation for the Chern-Simons current 
\begin{equation}
\delta K_{\mu} = \varepsilon_{\mu\nu}\partial^{\nu}\delta\Omega - \partial_{\mu}\delta\sigma
\end{equation}
We start with a uniform brane along the y-axis corresponding to a gauge configuration
\begin{equation} \label{inbrane}
A_x=0,\;\;A_y=2\pi\delta(x)
\end{equation}
which is obtained from (\ref{2dA}) by choosing 
\begin{equation} \label{flatbrane}
\sigma = 2\pi\Theta(x),\;\;\Omega = 0
\end{equation}
where $\Theta(x)$ is a unit step function. Let's now consider a combination of a physical variation and a gauge transformation
\begin{equation} \label{simgaOmega}
\delta\sigma = 2\pi\epsilon\Theta(x), ~~~ \delta\Omega = -2\pi\epsilon y \delta(x).
\end{equation}
It is easy to see that the in-brane component $A_y$ is invariant under (\ref{simgaOmega}), so is the field strength $F_{xy}$. However the transverse component becomes non-vanishing
\begin{equation} \label{Ax}
A_x = -2\pi\epsilon y \delta'(x).
\end{equation}
Therefore the variation (\ref{simgaOmega}) describes a brane fluctuation with respect to the original background (\ref{inbrane}). The physical field strength does not change but the transverse gauge potential develops a non-vanishing component which indicates the fluctuation of the brane. 
This leads to a variation of the Chern-Simons vector that can be interpreted as an infinitesimal spacetime
rotation of the original brane configuration (\ref{inbrane})
\begin{eqnarray} \label{deltaKxKy} 
\delta K_x & = & 2\pi\epsilon\delta(x) - 2\pi\epsilon\delta(x) = 0 \cr
\delta K_y & = & -2\pi\epsilon y \delta'(x)\approx 2\pi\left[\delta(x-\epsilon y) - \delta(x)\right]
\end{eqnarray}
This is exactly what we expected from the anomaly-inflow constraint (\ref{2DAI}) since (\ref{Ax}) and (\ref{deltaKxKy}) describe an infinitesimal rotation of the membrane (recall that $\partial_{\mu} \theta$ is proportional to the normal vector $n_{\mu}$ of the membrane). This can also be seen from the constraint $K_{\mu} A^{\mu} = 0 $. If the membrane rotates the gauge configuration $A_{\mu}$ also rotates in spacetime, the Chern-Simons current has to rotate as well to be perpendicular to $A_{\mu}$ (note this argument only works in two dimensions). The anomaly-inflow constraint then requires that $\partial_{\mu} \theta$ should rotate to be consistent with the rotation of the normal vector $n_{\mu}$ on the membrane.    

Similar to the string case, we can construct a topological source term for the membrane
\begin{equation} \label{2Dbrane}
S_{\theta} = \int d^2 x ~ \partial_{\mu} \theta \, K^{\mu} 
\end{equation}
which is equivalent to a two-dimensional topological charge term after integrating by parts
\begin{equation} \label{2Dtc}
S'_{\theta} = \int d^2x ~ \theta(x) \epsilon_{\mu\nu} F^{\mu\nu} 
\end{equation}
Behind this seeming equivalence, there is a significant difference between (\ref{2Dbrane}) and (\ref{2Dtc}). Usually the topological charge term (\ref{2Dtc}) is considered as gauge-invariant, while the topological current term (\ref{2Dbrane}) is not and its gauge variation is needed for cancelling the gauge anomaly localized on the lower dimensional topological defects.

\section{Discussions}

The membranes and the Chern-Simons current might be relevant to the confinement problem as well. In fact it has been realized by M. L$\ddot{u}$scher \cite{Luscher78} long time ago that the Chern-Simons current $K_{\mu}$ has a long range character, i.e. its propagator has a pole at $p^2=0$. However it does not correspond to a massless particle, similar to the gauge potential $A_{\mu}$ in two-dimensional QED. The two-point connected correlation function of the Chern-Simons current $K_{\mu}$ is \cite{Luscher78}
\begin{equation}
\langle
K_{\mu}(x) K_{\nu}(y)\rangle_{\theta} =\int \frac{d^4p}{(2\pi)^4} e^{i p(x-y)}\frac{P_{\mu} P_{\nu}}{p^2} G_{\theta} (p^2) + \cdots
\end{equation}
where the propagator has a pole at $p^2 =0$, as it can be seen from
\begin{equation} \label{pole}
G_{\theta} (p^2) = \frac{4\pi^4}{p^2} \frac{dF(\theta)}{d\theta},~(p^2 \rightarrow 0),~~F(\theta) = -i\langle Q_{TC}\rangle_{\theta}.
\end{equation}    
where $Q_{TC}$ is the topological charge density. In Ref.\cite{TX1} we have shown that the pole due to the massless Ramond-Ramond field cancels the "wrong sign" pole in (\ref{pole}) which leads to the Witten-Veneziano formula for topological susceptibility and the $\eta'$ mass term.
Here we consider the confinement problem as in the two-dimensional $CP^{(n-1)}$ model in Ref.\cite{Luscher78}. 
Note that the Wilson loop can be considered as a closed "membrane" in two dimensions
\begin{equation}
\exp \big( i e \oint_{C} dx^{\mu} A_{\mu} \big)
\end{equation}
and $A_{\mu}$ plays the role of Chern-Simons current in the two-dimensional $CP^{(n-1)}$ model. It has been shown \cite{Luscher78}
\begin{equation}
\langle \exp \big(i e \oint_{C} dx^{\mu} A_{\mu} \big)\rangle|_{\theta=0}  \propto \exp \{ - V \int^{2\pi e}_{0} d\theta F(\theta) \}, ~V \rightarrow \infty
\end{equation}
which leads to the area law and hence the confinement (when $e$ is not integral). In four dimensions one may consider the Chern-Simons current flowing into a three-dimensional membrane $\Sigma$ and expect that
\begin{equation}
\langle \exp \big( i \oint_{\Sigma} d\sigma^{\mu} K_{\mu} \big) \rangle|_{\theta=0} \propto \exp \{ - V \int d\theta F(\theta) \}, ~V \rightarrow \infty ,
\end{equation} 
however further investigations are needed to support these ideas.

To summarize, we have studied the membrane-like distribution of the topological charge and its anomaly inflow through the "topological currents", e.g. the Chern-Simons current in the QCD vacuum. The current-current coupling form, e.g. $\partial_{\mu} \theta K^{\mu}$, seems to be more appropriate than the usual topological charge term in describing extended topological defects such as vortices and membranes, especially when quarks and their currents are taken into account. This is because the quark zero modes localized on the topological defects usually have gauge anomaly, which can only be cancelled by the inflow of some topological current from the bulk space. In the membrane case, it is also suggested \cite{TX2} that near-zero quark modes may be localized on opposite sides of the membrane and this leads to a new scenario to study the chiral condensate. When the membrane fluctuates they may puncture through it and annihilate, which naturally leads to the U(1) axial anomaly and the mass insertion of $\eta'$. Topological currents like the Chern-Simons current could have long-range features which may be applied to the confinement problem.

\acknowledgments

This work was supported by the Department of Energy under grant DE-FG02-97ER41027. CX is supported by the research funds from the Institute of Advanced Studies, Nanyang Technological University, Singapore.


\begin {thebibliography}{99}

\bibitem{Witten79}
E.~Witten, Nucl. Phys. B149: 285 (1979).

\bibitem{Witten98} 
E.~Witten, Phys.~Rev.~Lett. 81: 2862 (1998).


\bibitem{Horvath03} 
  I.~Horvath, S.~J.~Dong, T.~Draper, F.~X.~Lee, K.~F.~Liu, N.~Mathur, H.~B.~Thacker and J.~B.~Zhang,
  %"Low dimensional long range topological charge structure in the QCD vacuum,''
  Phys.\ Rev.\ D {\bf 68}, 114505 (2003)
  [hep-lat/0302009].

\bibitem{Horvath05} 
  I.~Horvath, A.~Alexandru, J.~B.~Zhang, Y.~Chen, S.~J.~Dong, T.~Draper, F.~X.~Lee and K.~F.~Liu {\it et al.},
  %"Inherently global nature of topological charge fluctuations in QCD,''
  Phys.\ Lett.\ B {\bf 612}, 21 (2005)
  [hep-lat/0501025].

\bibitem{Ilgenfritz07}
  E.~-M.~Ilgenfritz, K.~Koller, Y.~Koma, G.~Schierholz, T.~Streuer and V.~Weinberg,
  %"Exploring the structure of the quenched QCD vacuum with overlap fermions,''
  Phys.\ Rev.\ D {\bf 76} (2007) 034506
  [arXiv:0705.0018 [hep-lat]].

\bibitem{Ilgenfritz08} 
  E.~-M.~Ilgenfritz, D.~Leinweber, P.~Moran, K.~Koller, G.~Schierholz and V.~Weinberg,
  %"Vacuum structure revealed by over-improved stout-link smearing compared with the overlap analysis for quenched QCD,''
  Phys.\ Rev.\ D {\bf 77}, 074502 (2008)
  [Erratum-ibid.\ D {\bf 77}, 099902 (2008)]
  [arXiv:0801.1725 [hep-lat]].

\bibitem{Hynes08} 
  P.~Keith-Hynes and H.~B.~Thacker,
  %"Fractionally charged Wilson loops as a probe of theta-dependence in CP(N-1) sigma models: Instantons vs. large N,''
  Phys.\ Rev.\ D {\bf 78}, 025009 (2008)
  [arXiv:0804.1534 [hep-lat]].

\bibitem{Thacker10}
H. Thacker, Phys.~Rev. D81, 125006 (2010).

\bibitem{TX1} 
  H.~B.~Thacker, C.~Xiong and A.~Kamat,
  %"Chiral quark dynamics and topological charge: The role of the Ramond-Ramond U(1) Gauge Field in Holographic QCD,''
  Phys.\ Rev.\ D {\bf 84}, 105011 (2011)
  [arXiv:1104.3063 [hep-th]].

\bibitem{Buividovich11} 
  P.~V.~Buividovich, T.~Kalaydzhyan and M.~I.~Polikarpov,
  %"Fractal dimension of the topological charge density distribution in SU(2) lattice gluodynamics,''
  Phys.\ Rev.\ D {\bf 86}, 074511 (2012)
  [arXiv:1111.6733 [hep-lat]].

\bibitem{Callan-Harvey}
C. Callan and J. Harvey, Nucl.~Phys. B250, 427 (1985).

\bibitem{Green}
M. Green, J. Harvey, and G. Moore, Class. Quant. Grav. 14, 47 (1997).

\bibitem{TX2} 
  H.~B.~Thacker and C.~Xiong,
  %"Anomaly Inflow and Membrane Dynamics in the QCD Vacuum,''  
  Phys.\ Rev.\ D {\bf 86}, 105020 (2012)  [arXiv:1208.4784 [hep-th]].  %%CITATION = ARXIV:1208.4784;%%

\bibitem{Clark} 
  T.~E.~Clark, S.~T.~Love, M.~Nitta, T.~ter Veldhuis and C.~Xiong,
  %"Brane Vector Dynamics from Embedding Geometry,''
  Nucl.\ Phys.\ B {\bf 810}, 97 (2009)
  [arXiv:0809.1083 [hep-th]].

\bibitem{Zumino83} 
  B.~Zumino,
  "Chiral Anomalies And Differential Geometry: Lectures Given At Les Houches, August 1983,''
  In *Treiman, S.b. ( Ed.) Et Al.: Current Algebra and Anomalies*, 361-391 and Lawrence Berkeley Lab. - LBL-16747 (83,REC.OCT.) 46p

\bibitem{Stora83} 
 R.~Stora,
  "Algebraic Structure And Topological Origin Of Anomalies,''
  LAPP-TH-94.
  %%CITATION = LAPP-TH-94;%% 


\bibitem{Kogut75}
J. Kogut and L. Susskind, Phys. Rev. D11, 3594 (1975).

\bibitem{Coleman75} 
  S.~R.~Coleman,
  %"More About the Massive Schwinger Model,''
  Annals Phys.\  {\bf 101}, 239 (1976).

\bibitem{Luscher78}
M. Luscher, Phys.~Lett. 78B, 465 (1978).
 
\end{thebibliography}
\end{document}